\documentclass[11pt]{article} 
\usepackage{latexsym}
\usepackage{amssymb,amsmath,amsfonts}
\usepackage{verbatim}
\usepackage{changebar}
\usepackage{setspace}
\usepackage{simplewick}
\usepackage{tocvsec2}
\usepackage{array}
\usepackage{cite}
\usepackage{mathtext}
\usepackage{stackrel}

\usepackage[usenames,dvipsnames]{xcolor}

\usepackage[colorlinks,hyperindex, 
            bookmarks=true,bookmarksopen=true]{hyperref}
    \hypersetup
    {
        colorlinks=true,
        linkcolor=MidnightBlue,
        urlcolor=MidnightBlue,
        filecolor=black,
        citecolor=RedOrange,
        pdfstartview=FitV,
        pdftitle={},
        pdfauthor={Ilmar Gahramanov},
        pdfsubject={},
        pdfkeywords={},
        pdfpagemode=None,
        bookmarksopen=true
    }

\begin{document}

\rightline{\vbox{\small\hbox{\tt HU-EP-13-44} }}
\vskip 2.7 cm

\centerline{\Large \bf A new pentagon identity for the tetrahedron index}
\vskip 1 cm

\centerline{\large {\bf Ilmar Gahramanov$^{a,b}$\footnote{\href{mailto:ilmar.gahramanov@physik.hu-berlin.de}{{\textbf{ilmar.gahramanov@physik.hu-berlin.de}}}} and Hjalmar Rosengren$^{c}$\footnote{\href{mailto:hjalmar@chalmers.se}{\textbf{hjalmar@chalmers.se}},  \href{http://www.math.chalmers.se/~hjalmar/}{http://www.math.chalmers.se/{\textasciitilde}hjalmar} }}  }

\begin{center}
\textit{ $^a$Institut f\"{u}r Physik, Humboldt-Universit\"{a}t zu Berlin,\\ 
Newtonstrasse 15, 12489 Berlin, Germany} \\
\texttt{} \\
\vspace{.2mm}
\textit{ $^b$Institute of Radiation Problems ANAS,\\ B.Vahabzade 9, AZ1143 Baku, Azerbaijan} \\
\texttt{} \\
\vspace{.2mm}
\textit{ $^c$Department of Mathematical Sciences \\ Chalmers University of Technology and University of Gothenburg\\SE-412~96 G\"{o}teborg, Sweden} \\
\texttt{} \\
\vspace{.1mm}
\end{center}

\vskip 1.5cm \centerline{\bf Abstract} \vskip 0.2cm \noindent Recently Kashaev, Luo and Vartanov, using the reduction from a four--dimensional superconformal index to a three--dimensional partition function, found a pentagon identity for a special combination of hyperbolic Gamma functions. Following their idea we have obtained a new pentagon identity for a certain combination of so--called tetrahedron indices arising from the equality of  superconformal indices of dual three--dimensional ${\cal N}=2$ supersymmetric theories and give a mathematical proof of it.


\section{Introduction}

One of the efficient tools in the study of supersymmetric gauge theories is the superconformal index. It was introduced in  \cite{Romelsberger:2005eg, Kinney:2005ej, Romelsberger:2007ec} and it counts the number of gauge invariant protected short operators in the theory. The interest in this quantity is motivated by various applications, suffice it to say that the superconformal index technique provides the most rigorous mathematical check of Seiberg and Seiberg--like dualities.  

The superconformal index of a four--dimensional supersymmetric gauge theory is expressed in terms of elliptic hypergeometric integrals. The identification of superconformal indices of Seiberg dual theories is nothing but Weyl group symmetry transformations for certain elliptic hypergeometric functions.  The relation between superconformal indices and elliptic hypergeometric integrals (see, for instance, \cite{Dolan:2008qi, Spiridonov:2008zr, Gadde:2009kb, Spiridonov:2009za,  Gadde:2010te, Spiridonov:2010hh, Spiridonov:2010qv, Vartanov:2010xj, Spiridonov:2011hf, Teschner:2012em, Spiridonov:2012ww, Dimofte:2012pd, GarciaEtxebarria:2012qx, Spiridonov:2012de, Gahramanov:gka, Gahramanov:2013xsa}) allows for checks of known supersymmetric dualities and investigations of new ones, and also the discovery of complicated new integral identities. 

Another class of special functions appears in the calculation of superconformal indices for three--dimensional supersymmetric field theories. The three--dimensional index can be expressed in terms of the so--called tetrahedron index. In the present paper, we have obtained an interesting mathematical identity for this function from the mirror duality of $3d$ ${\cal N}=2$ theories. 

The paper is organized in the following way. In Section 2 we make a brief review of the superconformal index and we discuss $4d \rightarrow 3d$ reduction and the pentagon identity of a certain combination of hyperbolic Gamma functions. We describe our result in Section 3 and briefly discuss some open problems. In Section 4 we give a mathematical proof of the new pentagon identity. The following ``commutative diagram'' demonstrates the plan of the paper pictorially (we use the shorthands SCI -- superconformal index and PF -- partition function). 

\vspace{-0.6cm}

{\small
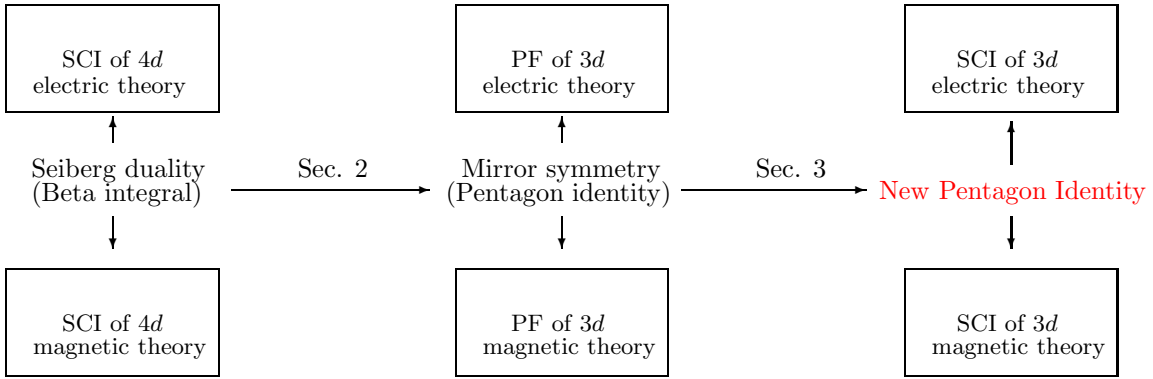
\begin{picture}(100,170)(-10,5)
\put(0,0){\framebox(80,40){{\footnotesize SCI of $4d$}}}
\put(10,8){{\footnotesize  magnetic theory}}
\put(0,100){\framebox(80,40){{\footnotesize SCI of $4d$}}}
\put(10,107){{\footnotesize  electric theory}}
\put(170,0){\framebox(80,40){{\footnotesize PF of $3d$  }}}
\put(180,8){{\footnotesize magnetic theory}}
\put(170,100){\framebox(80,40){{\footnotesize PF of $3d$ }}}
\put(180,107){{\footnotesize electric theory}}
\put(340,0){\framebox(80,40){{\footnotesize SCI of $3d$ }}}
\put(350,8){{\footnotesize magnetic theory}}
\put(340,100){\framebox(80,40){{\footnotesize SCI of $3d$ }}}
\put(350,107){{\footnotesize electric theory}}
\put(85,70){\vector(1,0){75}}
\put(110,75){Sec. 2}
\put(40,88){{\vector(0,1){10}}}
\put(40,60){{\vector(0,-1){12}}}
\put(6,75){ Seiberg duality}
\put(6,66){ (Beta integral)}
\put(210,88){{\vector(0,1){10}}}
\put(210,60){{\vector(0,-1){12}}}
\put(172,75){Mirror symmetry}
\put(166,66){(Pentagon identity)}
\put(255,70){\vector(1,0){70}}
\put(280,75){ Sec. 3}
\put(330,67){{\color{red}New Pentagon Identity}}
\put(380,80){{\vector(0,1){16}}}
\put(380,60){{\vector(0,-1){12}}}

\end{picture}}

\vspace{0.5cm}

\section{A review of the index and the pentagon identities}

This section provides a very short review of the superconformal index for four-- and three--dimensional supersymmetric gauge theories. More details and references can be found, for instance, in \cite{Spiridonov:2009za, Sudano:2011aa, Gahramanov:gka} for $4d$ and in \cite{Bhattacharya:2008bja, Krattenthaler:2011da, Hwang:2011qt, Imamura:2011su, Yokoyama:2011qu} for $3d$ theories. The equality of the superconformal index of dual theories leads to non--trivial mathematical identities, some of which we will discuss.

\subsection{A few words about superconformal index}

The superconformal index for a ${\cal N}=1$ $d=4$ superconformal field theory is a non--trivial extension of the Witten index
\begin{equation}\label{Ind}
{\rm ind}(p,q,\underline{g},\underline{f}) \ = \ \text{Tr}  (-1)^{\cal F} e^{-\beta {\cal H} } p^{\mathcal{R}/2+J_3}q^{\mathcal{R}/2-J_3} e^{\sum_i f_i F_i} e^{\sum_j g_j G_j} \;,
\end{equation}
where $\mathcal{R}=R+2J_3$ and the trace is taken over the states satisfying\footnote{The index is defined in the radial quantization, therefore $Q^{\dagger}=S$.}
\begin{align} 
{\cal H}  = \frac12 \{\bar{Q}_1, -\bar{S}^1\} = H-2\bar{J}_3-\frac32 R = 0 \; .
\end{align}
Here ${\cal F}$ is the fermion number, $H$, $J_3$, $R$ are Cartans of the superconformal group and $p$, $q$, $g_k$ (gauge group), $f_k$ (flavor group) are the group parameters (chemical potentials) with maximal torus generators $F_i$ and $G_j$, respectively.  In order to count gauge invariant operators we integrate over the gauge group and refine the index\footnote{In other words, to get gauge singlets we impose the Gauss law constraint. For this reason the index is independent of the gauge moduli.}
\begin{equation}
I(p,q,\underline{t}) \ = \ \oint [dg] \; \; {\rm ind}(p,q,\underline{z},\underline{t}),
\end{equation}
where $[dg]$ is the invariant Haar measure. Here we have used fugacities $z_i=e^{g_i}$ and $t_i=e^{f_i}$ instead of the chemical potentials $g_i$ and $f_i$.

Dolan and Osborn observed \cite{Dolan:2008qi} that the superconformal index of $4d$ ${\cal N}=1$ theories is expressible in terms of integrals over elliptic Gamma functions (the elliptic hypergeometric integrals). The elliptic Gamma function was introduced by Ruijsenaars \cite{Ruijsenaars:1997:FOA} and is defined as
\begin{equation}
\Gamma(z,p,q) = \prod_{i,j=0}^{\infty} \frac{1-z^{-1} p^{i+1} q^{j+1}}{1-z p^i q^j}, \; \; ~~~~ \text{with $|p|,|q|<1$.}
\end{equation}
The elliptic hypergeometric integrals are a new class of functions discovered by Spiridonov \cite{Spiridonovbeta, SpiridonovV}. We refer to the review \cite{Spiridonovessay} for details of such integrals.

We will use the $4d$ index in subsection $2.3$, however, more important for our present purpose is the superconformal index of three--dimensional supersymmetric gauge theories. The index for a ${\cal N}=2$ $d=3$ superconformal field theory is defined \cite{Bhattacharya:2008bja} in a similar way to (\ref{Ind}) 
\begin{equation} \label{3dindex}
{\rm ind}(q,\underline{t},\underline{z}) \ = \ \text{Tr} (-1)^F e^{-\beta {\cal H}} q^{\frac12 (\Delta+J_3)} \prod_{i=1}t_i^{F_i} \prod_{j=1}z_j^{G_j}
\end{equation}
where the trace is taken over the Hilbert space of the theory on a two--sphere $S^2$; $\Delta$, $J_3$ and $R$ are Cartans of the superconformal group and the fugacities $t_i$ and $z_j$ are associated with the flavor and gauge groups, respectively. Similarly to the $4d$ case to get the full index one needs to integrate the last expression over the gauge group.

The index of a $3d$ ${\cal N}=2$ theory can be written in terms of the so-called tetrahedron index. The tetrahedron index is defined by the following expression \cite{Dimofte:2011ju, Dimofte:2011py}
\begin{equation}
{\cal I}_q[m,z] \ = \ \prod_{i=0}^{\infty} \frac{1-q^{i-\frac12 m+1}z^{-1}}{1-q^{i-\frac12 m} z}, \;\;\; \text{with $|q|<1$ and $m \in Z$.}
\end{equation}
This is an index of free chiral multiplet with zero R-charge. The integer parameter $m$ stands for the magnetic charge. The matter of fact that in the definition (\ref{3dindex}) we implicitly sum over magnetic fluxes on $S^2$ which appears in the localization procedure as a contribution of monopoles. We take into account the summation by labeling the Hilbert space of the theory with magnetic charge $m$. 

To our knowledge, the tetrahedron index does not appear to be well known in the mathematical community; there are only a few papers \cite{Garoufalidis3d, Garoufalidis3d2}\footnote{Notice that in \cite{Garoufalidis3d, Garoufalidis3d2} the tetrahedron index is a Fourier coefficient of the tetrahedron index mentioned here.} on this subject. It is worth mentioning here that the tetrahedron index is a $q$-series version of the quantum dilogarithm of Faddeev and Kashaev \cite{Faddeev:1993rs, Kashaev:1994pj, Kashaev:1996kc}.

Here we have given definitions of the superconformal index for ${\cal N}=1$ and ${\cal N}=2$ theories. In fact, this quantity can be defined for any supersymmetric gauge theory \cite{Kinney:2005ej}, even for non--relativistic superconformal field theories \cite{Nakayama:2008qm}.

\subsection{The first pentagon identity for the tetrahedron index}

Let us consider the ${\cal N}=2$ $d=3$ supersymmetric field theory with $U(1)$ gauge group and one flavor\footnote{In this subsection we follow \cite{Krattenthaler:2011da}.}. The superconformal index of this theory is
\begin{equation}  \label{mirrorsym1}
I_e \ = \ \sum_{m \in Z} q^{|m|/3} \int_{\mathbb{T}} \frac{dz}{2\pi i z} \frac{(q^{5/6+|m|/2}z;q)_{\infty}(q^{5/6+|m|/2}z^{-1};q)_{\infty}}{(q^{1/6+|m|/2}z;q)_{\infty}(q^{1/6+|m|/2}z^{-1};q)_{\infty}}
\end{equation}
where the q-Pochhammer symbol is defined as $(z;q)_\infty=\prod_{i=0}^\infty (1-z q^i)$ and $\mathbb{T}$ denotes the unit circle with positive orientation. We bypass the influence of the topological symmetry $U(1)_J$ to the index. We refer the reader to \cite{Krattenthaler:2011da} for the expression in the presence of this symmetry. 

The mirror partner of this theory is the free Wess--Zumino theory \cite{Intriligator:1996ex, deBoer:1997ka, Aharony:1997bx}. The index of the Wess--Zumino theory is given by the simpler expression
\begin{equation} 
I_m \ = \ \frac{(q^{2/3};q)^3_{\infty}}{(q^{1/3};q)^3_{\infty}}
\end{equation}

As a result of mirror symmetry the indices of these two theories should be equal\footnote{For the proof see \cite{Krattenthaler:2011da}, also see \cite{Imamura:2011su} for a check via series expansion.}
\begin{equation} \label{mirrorsym2}
I_e \ = \ I_m \;.
\end{equation}
One can rewrite these indices in terms of tetrahedron indices. Then the equality of indices (\ref{mirrorsym2}) turns out to be the pentagon identity for the tetrahedron index
\begin{equation} \label{firstpentagon}
\boxed{ \sum_{m \in Z} \oint \frac{dz}{2\pi i z} z^{-m} \; {\cal I}_q[m;q^{1/6}z^{-1}] \; {\cal I}_q[-m;q^{1/6}z] \ = \ {\cal I}^3_q[0;q^{1/3}]. }
\end{equation}
This is the first example of a non--trivial identity for the tetrahedron index. Note that to obtain the left side of (\ref{firstpentagon}) we have used the following identity \cite{Dimofte:2011py}
\begin{equation}
(-q^{\frac12})^{\frac12(m+|m|)} z^{-\frac12(m+|m|)} \prod_{i=0}^{\infty} \frac{1-q^{i+\frac12 |m|+1} z^{-1}} {1-q^{i+\frac12|m| }z} \ = \ {\cal I}_q[m,z].
\end{equation}

\subsection{3d reduction}

In this subsection we discuss some aspects of \cite{Kashaev:2012cz} which are useful for the considerations in this paper. Let us consider the elliptic beta integral\footnote{Interestingly, this integral identity arises in different fields of theoretical physics, particularly, it is a star--triangle relation of an integrable lattice model \cite{Bazhanov:2010kz,Bazhanov:2011mz}. Also note that limits of the beta integral lead to many identities for hypergeometric integrals, for instance, the limit $p \rightarrow 0$ gives the Nassrallah--Rahman integral \cite{Nassrallah}. Other physical applications are discussed in \cite{Spiridonov:2011hf, Spiridonov:2010em, Spiridonov:2013zma}.} \cite{Spiridonovbeta} 
\begin{equation} \label{Betaint}
\frac{(p;p)_\infty (q;q)_\infty}{2} \int_{\mathbb{T}} \frac{\prod_{i=1}^6 \Gamma(t_i z ;p,q)\Gamma(t_i z^{-1} ;p,q)}{\Gamma(z^{2};p,q) \Gamma(z^{-2};p,q)} \frac{dz}{2 \pi \textup{i} z} = \prod_{1 \leq i < j \leq 6} \Gamma(t_i t_j;p,q),
\end{equation}
where $t_j,\; j=1,\ldots,6$ are complex parameters with the balancing condition $\prod_{j=1}^6 t_j=pq$. From the physical point of view the integral on the left hand side of the expression (\ref{Betaint})  is the index of the $4d$ ${\cal N}=1$ electric theory with $SU(2)$ gauge group and $N_F=3$ flavors, chiral scalar multiplets in the fundamental representation of the flavor group, while the expression on the right side is for the dual magnetic theory with chirals in the antisymmetric tensor representation of the flavor group. 

There is a nice reduction procedure from the $4d$ index to the $3d$ partition function proposed by Dolan et al. in \cite{Dolan:2011rp} (see also \cite{Gadde:2011ia,Imamura:2011uw}). A compelling physical argument for this reduction has been provided in \cite{Aharony:2013dha} (see also \cite{Niarchos:2012ah}). The essential step in the reduction scheme is scaling the fugacities in the following way
\begin{equation}
p=e^{2\pi i v \omega_1}, \;\; q=e^{2 \pi i v \omega_2}, \;\; z=e^{2 \pi i v u}, \;\; t_i=e^{2\pi i v \alpha_i} \; .
\end{equation}
Then the $3d$ partition function can be achieved by taking $v \rightarrow 0$ limit of the $4d$ superconformal index and by integrating out massive fields.  Geometrically, we consider a four--dimensional superconformal field theory on $S^3 \times S^1$, the limit $v \rightarrow 0$  shrinks the $S^1$ to zero and gives rise to a field theory on a squashed $S^3$. From the perspective of special functions this reduction brings elliptic Gamma functions to hyperbolic Gamma functions. Using properties of elliptic Gamma functions and hyperbolic Gamma functions (for details, see \cite{Kashaev:2012cz}), it is straightforward to derive the partition function from the elliptic beta integral via the above recipe
\begin{equation} \label{indexrel}
\int_{-\textup{i} \infty}^{\textup{i} \infty} \prod_{i=1}^3 \gamma^{(2)}(a_i - u;\omega_1,\omega_2) \gamma^{(2)}(b_i + u;\omega_1,\omega_2) \frac{du}{\textup{i} \sqrt{\omega_1\omega_2}} = \prod_{i,j=1}^3 \gamma^{(2)}(a_i+b_j;\omega_1,\omega_2),
\end{equation}
with the balancing condition $\sum_{i=1}^3 (a_i+b_i) = \omega_1+\omega_2$. Here the function $\gamma^{2}(u; \omega_1, \omega_2)$ is a hyperbolic Gamma function
\begin{equation}
\gamma^{(2)}(u;\omega_1,\omega_2) = e^{-\pi \textup{i}
B_{2,2}(u;\omega_1,\omega_2)/2} \frac{(e^{2 \pi i u/\omega_1}
\widetilde{q};\widetilde{q})_\infty}{(e^{2 \pi i
u/\omega_2};q)_\infty},
\end{equation}
where
\begin{equation}
 q = e^{2 \pi i \omega_1/\omega_2},\qquad  \widetilde{q} = e^{-2 \pi i
\omega_2/\omega_1},
\end{equation} 
and $B_{2,2}(u;\omega_1,\omega_2)$ denotes the second order Bernoulli polynomial, 
\begin{equation}
 B_{2,2}(u;\omega_1,\omega_2) =
\frac{u^2}{\omega_1\omega_2} - \frac{u}{\omega_1} -
\frac{u}{\omega_2} + \frac{\omega_1}{6\omega_2} +
\frac{\omega_2}{6\omega_1} + \frac 12.
\end{equation}
There are different notations and modifications of hyperbolic Gamma function, relations between some of them can be found in \cite{Galakhov:2012hy, Spiridonov:2010em} (also see the appendix of \cite{Hadasz:2013bwa}). 

Let us introduce the following function
\begin{equation} \label{BB}
\mathcal{B}(x,y) =
\frac{\gamma^{(2)}(x;\omega_1,\omega_2)\gamma^{(2)}(y;\omega_1,\omega_2)}{\gamma^{(2)}(x+y;\omega_1,\omega_2)}.
\end{equation}
Then from the expression (\ref{indexrel}) one can easily see that this function satisfies the pentagon identity \cite{Kashaev:2012cz}
\begin{equation} \label{Kashaevpent}
\int_{-\textup{i} \infty}^{\textup{i} \infty} \prod_{i=1}^3
\mathcal{B}(a_i - u,b_i + u) \frac{du}{\textup{i}
\sqrt{\omega_1\omega_2}} = \mathcal{B}(a_2+b_1,a_3+b_2)
\mathcal{B}(a_1+b_2,a_3+b_1). 
\end{equation}

\section{The second Pentagon identity for the tetrahedron index}

Motivated by the result (\ref{Kashaevpent}) obtained by Kashaev et al. in \cite{Kashaev:2012cz}, we wonder whether a similar identity arises on the level of superconformal indices.

In fact without any computations one can also get the result for the three--dimensional superconformal index by using the relation between a partition function and an index. To obtain the superconformal index from the partition function of a $3d$ ${\cal N}=2$ theory, roughly speaking, one should change all hyperbolic Gamma functions to a tetrahedron index
\begin{equation}
\gamma^{(2)}(a \pm b;\omega_1, \omega_2) \quad \rightarrow \quad {\cal I}_q [m; a \; b^{\pm 1}]
\end{equation}
To our knowledge there is no rigorous mathematical proof of this relation. However there are a lot of examples where this relation works perfectly. For instance, the pentagon identity corresponding to the expression (\ref{firstpentagon}) was found in \cite{Dimofte:2011ju} in terms of hyperbolic Gamma functions. Also one can observe this relation by comparing the results of \cite{Aharony:2013dha, Gahramanov:gka} and \cite{Park:2013wta}.  

Using this correspondence one can construct from (\ref{indexrel}) the following relation for the indices of the theories in question
\begin{equation} \label{prepentagon}
\sum_{m \in Z} \oint \frac{d z}{2 \pi i z} \; (-z)^{-3 m} \prod_{i=1}^{3} {\cal I}_q[-m,q^{\frac16}\xi_i z] \; {\cal I}_q[m,q^{\frac16}\eta_i z^{-1}] = \prod_{i,j=1}^3 {\cal I}_q[0,q^{\frac13}\xi_i \eta_j] \; ,
\end{equation}
where $\prod_{i=1}^3 \xi_i = \prod_{i=1}^3  \eta_i =1$ which is the analogue of the balancing condition for the expression (\ref{indexrel}). This expression is a consequence of a mirror symmetry. On the left side we have the $3d$ ${\cal N}=2$ superconformal field theory with $U(1)$ gauge symmetry and six chiral multiplets, while the mirror partner on the right side has nine chirals \footnote{We again dropped the topological symmetry $U(1)_J$.}. We have, experimentally, checked the expression (\ref{prepentagon}) by expansion of  each side of the equality in powers of $q$ and setting all chemical potentials $\xi_i$ and $\eta_i$ to $1$ and obtained the same result 
\begin{equation}
1+9 q^{1/3}+36 q^{2/3}+84 q+135 q^{4/3}+198 q^{5/3}+327 q^2+477 q^{7/3}+504 q^{8/3}+568 q^3+\dotsm.
\end{equation}
Let us introduce the following function analogous to (\ref{BB})
\begin{equation}
{\cal B}[m;a,b]=\frac{{\cal I}_q[m,a] \; {\cal I}_q[-m,b]}{{\cal I}_q[0,a b]} \; ,
\end{equation}
and rewrite the expression (\ref{prepentagon}) in terms of this function. The final result is
\begin{equation} \label{pentagon}
\boxed{ \sum_{m\in Z} \oint \frac{d z}{2 \pi i z}  \; (-z)^{-3 m} \prod_{i=1}^3 {\cal B}[m; \xi_i z^{-1}, \eta_i z]= {\cal B}[0; \xi_1 \eta_2, \xi_3 \eta_1] \; {\cal B}[0; \xi_2 \eta_1,  \xi_3 \eta_2]}
\end{equation}
where we have redefined the fugacities  $\xi_i \rightarrow q^{-1/6} \xi_i$ and $\eta_i \rightarrow q^{-1/6} \eta_i$. Then the new  balancing condition is $\prod_{i=1}^3 \xi_i = \prod_{i=1}^3  \eta_i =q$.  Notice that, in deriving the product of two ${\cal B}[m;a,b]$ functions on the right side of (\ref{pentagon}) we have used the following property of the tetrahedron index
\begin{equation} \label{notmassive}
{\cal I}_q[m,a] \; {\cal I}_q[m, q a^{-1}]=1 \;,
\end{equation} 
which is similar to the inversion relation for hyperbolic Gamma functions
\begin{equation}
\gamma^{(2)}(a;\omega_1,\omega_2) \; \gamma^{(2)}(\omega_1+\omega_2-a;\omega_1,\omega_2)=1.
\end{equation}
From physical point of view the property (\ref{notmassive}) of the tetrahedron index means that the massive fields do not contribute to the index of the theory.

As we see from above, the function ${\cal B}[m;a,b]$ satisfies the pentagon identity. This particular example clearly demonstrates the power of the superconformal index technique in gaining extremely nontrivial integral identities.

Let us conclude by making a brief comment on the geometrical interpretation of our result. There is a nice relation between $3d$ ${\cal N}=2$ supersymmetric gauge theories and $3$--manifolds known as ``class R'' \cite{Dimofte:2011ju} (see also \cite{Dimofte:2011py, Dimofte:2013iv}). In this context the pentagon identity (\ref{pentagon}) encodes information about the geometry of the corresponding 3--manifolds and it would be interesting to understand the meaning of the pentagon identity from this point of view.

\section{Proof of the second pentagon identity}

In this section, we will use the standard notation
$$(a;q)_k=\prod_{j=0}^{k-1}(1-aq^j),\qquad k=0,1,2,\dots,\infty, $$
$$(a_1,\dots,a_m;q)_k=(a_1;q)_k\dotsm (a_m;q)_k, $$
$${}_3\phi_2\left(\begin{matrix}a,b,c\\d,e\end{matrix};q,z\right)=\sum_{k=0}^\infty\frac{(a,b,c;q)_k}{(q,d,e;q)_k}\,z^k. $$

 Making the change of parameters
$\xi_j=q^{-1/6}a_j$, $\eta_j=q^{-1/6}b_j$, \eqref{prepentagon}
can be stated as
\begin{equation}\label{gc}
\sum_{m=-\infty}^\infty\oint\frac{dz}{2\pi i z} (-z)^{-3m}\prod_{i=1}^3\frac{(q^{1+\frac m2}/a _i z,q^{1-\frac m2}z/b_i;q)_\infty}{(q^{\frac m2}a_i z,q^{-\frac m2}b_i/z;q)_\infty}=\prod_{i,j=1}^3\frac{(q/a_ib_j;q)_\infty}{(a_ib_j;q)_\infty},
 \end{equation}
where $|q|<1$, the parameters satisfy 
\begin{equation}\label{wb}
a_1a_2a_3=b_1b_2b_3=q,
\end{equation}
and the integration is over any positively oriented contour separating the  sequences of poles tending to zero from those tending to infinity. 

To describe the location of the poles, note that the factor $(q^{-\frac m 2}b_1/z)_\infty$ vanishes at $z=q^{-m/2+k}b_1$ for $k$ a non-negative integer. However, for $k\leq m-1$, this apparent singularity is cancelled by the factor $(q^{1-\frac m2}z/b_1)_\infty$. A similar phenomenon holds for all  factors, and we conclude that the integrand in \eqref{gc} has poles at
\begin{align}\label{ip} z&=q^{-\frac m2+k}b_j,& 1&\leq j\leq 3, & k&\geq\max(0,m).\\
\label{up} z&=q^{-\frac m2-k}/a_j,& 1&\leq j\leq 3, & k&\geq\max(0,-m).
\end{align}
The contour of integration must  separate the points \eqref{ip} from the points \eqref{up}, which is  always possible for generic parameters subject to \eqref{wb}. For instance,  if $|a_j|,\,|b_j|<1$ for all $j$, we may integrate over the unit circle.

To prove \eqref{gc}, we shrink the contour to $0$, picking up residues at the points \eqref{ip}. The resulting residue sum is a special case of \cite[Eq.\ (4.10.5)]{gr}, except that the range of summation in the ${}_6\phi_5$ sum on the right must be restricted from $k\geq 0$ to $k\geq\max(0,m)$. After standard simplification, the left-hand side of \eqref{gc} takes the form
\begin{multline}\label{rs}\sum_{m=-\infty}^\infty\sum_{k=\max(0,m)}^\infty
\frac{(qb_1/b_2,qb_1/b_3,q/a_1b_1,q/a_2b_1,q/a_3b_1;q)_\infty}{(b_2/b_1,b_3/b_1,a_1b_1,a_2b_1,a_3b_1;q)_\infty} \\
\times\frac{(a_1b_1,a_2b_1,a_3b_1;q)_k}{(q,qb_1/b_2,qb_1/b_3;q)_k}\,q^k\frac{(a_1b_1,a_2b_1,a_3b_1;q)_{k-m}}{(q,qb_1/b_2,qb_1/b_3;q)_{k-m}}\,q^{k-m}
+\operatorname{idem}(b_1;b_2,b_3),
\end{multline}
where $\operatorname{idem}(b_1;b_2,b_3)$ denotes the sum of the two terms obtained from the first one by interchanging $b_1$ with $b_2$ and $b_3$, respectively. After replacing $k-m$ by $m$, we find that \eqref{gc} is equivalent to the hypergeometric identity
\begin{multline}\label{ssi}
\frac{(qb_1/b_2,qb_1/b_3,q/a_1b_1,q/a_2b_1,q/a_3b_1;q)_\infty}{(b_2/b_1,b_3/b_1,a_1b_1,a_2b_1,a_3b_1;q)_\infty}\, {}_3\phi_2\left(\begin{matrix}a_1b_1,a_2b_1,a_3b_1\\qb_1/b_2,qb_1/b_3\end{matrix};q,q\right)^2\\
+\operatorname{idem}(b_1;b_2,b_3)=\prod_{i,j=1}^3\frac{(q/a_ib_j;q)_\infty}{(a_ib_j;q)_\infty}.
\end{multline}

Although we have not found \eqref{ssi} in the literature,  it can be reduced to known results by elementary means. To prove it, let
$$x_1=\frac{b_1(qb_1/b_2,qb_1/b_3;q)_\infty}{(a_1b_1,a_2b_1,a_3b_1;q)_\infty}\, {}_3\phi_2\left(\begin{matrix}a_1b_1,a_2b_1,a_3b_1\\qb_1/b_2,qb_1/b_3\end{matrix};q,q\right) $$
and let $x_2$ and $x_3$ be the same expression with $b_1$ interchanged with $b_2$ and $b_3$, respectively. Then, by the non-terminating $q$-Saalsch\"utz summation \cite[Eq.\ (II.24)]{gr},
$$x_2-x_1=(b_2-b_1)\frac{(qb_1/b_2,qb_2/b_1,q/a_1b_3,q/a_2b_3,q/a_3b_3;q)_\infty}{(a_1b_1,a_2b_1,a_3b_1,a_1b_2,a_2b_2,a_3b_2;q)_\infty}. $$
By symmetry, similar identities hold for  $x_1-x_3$ and $x_3-x_2$. Inserting these expressions into the elementary identity
$$(x_3-x_2)x_1^2+(x_1-x_3)x_2^2+(x_2-x_1)x_3^2=(x_2-x_1)(x_3-x_2)(x_3-x_1) $$
 yields \eqref{ssi} after simplification. This completes the proof of \eqref{prepentagon}.

It would be interesting to find an extension of \eqref{prepentagon} related in a similar way to the nonterminating Jackson summation
\cite[Eq.\ (II.25)]{gr}, which is a natural generalization of the $q$-Saalsch\"utz summation to the level of ${}_8W_7$-series.

\vspace{0.5cm}

\noindent \textbf{Acknowledgments.} IG is grateful to Tudor Dimofte, Ludvig Faddeev, Davide Gaiotto and J\"{o}rg Teschner for valuable discussions and would like to specially thank Grigory Vartanov for enlightening discussions, for sharing his ideas and for initiating this project. IG is also grateful to Ben Hoare and Jan Plefka for valuable improvements on the manuscript. HR is supported by the Swedish Science Research Council.

\begin{singlespace}

\end{singlespace}

\end{document}